\documentclass[12pt,preprint]{aastex}

\shorttitle{The rotation of Io predicted by the Poincar\'e model}
\shortauthors{Noyelles}

\usepackage{amsmath}
\usepackage{wasysym}

\begin{document}


\title{The rotation of Io predicted by the Poincar\'e-Hough model}


\author{Beno\^it Noyelles\altaffilmark{1}}
\affil{NAmur Center for Complex SYStems (NAXYS), University of Namur, Dpt of Mathematics, 8 Rempart de la Vierge,
B-5000 Namur, Belgium}


\altaffiltext{1}{email: Benoit.Noyelles@fundp.ac.be, also associated with IMCCE, CNRS UMR 8028, Paris Observatory, UPMC, USTL, 77 avenue Denfert-Rochereau, 75014 Paris, France}


\begin{abstract}

	\par This note tackles the problem of the rotation of Io with the 4-degrees of 
	freedom Poincar\'e-Hough model. Io is modeled as a 2-layer body, i.e. a triaxial 
	fluid core and a rigid outer layer.
	
	\par We show that the longitudinal librations should have an amplitude of about 30 arcseconds,
	independent of the composition of the core. We also estimate the tidal instability of the core,
	and show that should be slowly unstable.

\end{abstract}





\section{Introduction}

\par Thanks to the Galileo space mission in the Jovian system, we have now interesting clues on Io's internal 
structure. In particular, its mass and its second-order gravity field coefficients $J_2$ and $C_{22}$ are known 
with a good accuracy \citep{ajlms2001}. Moreover, internal heating due to an intense tidal dissipation 
\citep{pcr1979,lakv2009} is expected to induce a fluid core \citep{cpr1982}.

\par We recently published a theoretical exploration of the Poincar\'e-Hough model applied to a synchronously 
rotating body \citep{n2012}. This model, originally proposed independently by \citet{h1895} and \citet{p1910} 
before being put in an Hamiltonian form by \citet{tw2001}, describes the rotational dynamics of a triaxial body 
composed of a rigid mantle and an ellipsoidal cavity filled by an inviscid fluid of constant uniform density and 
vorticity. This is a 4-degrees of freedom conservative model in which the core-mantle interactions result in 
pressure coupling at the core-mantle boundary.

\par This paper proposes an application of this model to a realistic Io under the gravitational forcing of 
Jupiter. We build internal models and apply the Poincar\'e-Hough model to them, and we use complete orbital 
ephemerides, here from \citep{ldv2006}, to model accurately the gravitational torque of Jupiter. After computation of Io's 
rotation we will discuss the problem of the elliptical instability of the fluid filling the core.

\section{The internal structure of Io}

\par We here consider Io to be an ellipsoidal body, its moments of inertia being $0<A\leq B\leq C$. $A$ is the moment of inertia with respect 
to Io's long equatorial axis, while $C$ is related to the polar axis. The core is ellipsoidal as well, its principal axes of inertia being 
collinear to the ones of the whole body, the moments associated being written $A_c$, $B_c$ and $C_c$. The known gravity and shape parameters
of Io are gathered in Tab.\ref{tab:param}.

\placetable{tab:param}

\begin{table}[ht]
	\caption{Gravity and shape parameters of Io. $I$ has not been measured independently of the 
others, it is deduced from the measured values of $J_2$ and $C_{22}$ combined with the hydrostatic equilibrium condition.\label{tab:param}}
\centering
\begin{tabular}{l|r|r}
Quantity & Value & Reference \\
\hline
Mean density $\bar{\rho}$ & $3,527.8\pm2.9$ $kg/m^3$ & \citet{ajlms2001}\\
$J_2$ & $(1.8459\pm0.0042)\times10^{-3}$ & \citet{ajlms2001}\\
$C_{22}$ & $(5.537\pm0.012)\times10^{-4}$ & \citet{ajlms2001}\\
$I/\left(MR^2\right)$ & $0.37685\pm0.00035$ & \citet{ajlms2001}\\
Mean radius R & $1,821.49$ km & \citet{aabcccfhhknossttw2011} \\
Subplanetary equatorial radius a & $1,829.4$ km & \citet{aabcccfhhknossttw2011} \\
Along orbit equatorial radius b & $1,819.4$ km & \citet{aabcccfhhknossttw2011} \\
Polar radius c & $1,815.7$ km & \citet{aabcccfhhknossttw2011}
\end{tabular}
\end{table}

The Poincar\'e-Hough rotational model requires 5 internal structure parameters:

\begin{itemize}

\item Polar flattening $\epsilon_1=\frac{2C-A-B}{2C}=J_2\frac{MR^2}{C}$,

\item Equatorial ellipticity $\epsilon_2=\frac{B-A}{2C}=2C_{22}\frac{MR^2}{C}$,

\item Polar flattening of the core $\epsilon_3=\frac{2C_c-A_c-B_c}{2C_c}$,

\item Equatorial ellipticity of the core $\epsilon_4=\frac{B_c-A_c}{2C_c}$,

\item Relative inertia of the core $\delta=\frac{C_c}{C}$.

\end{itemize}

\par From the definitions of $J_2=(2C-A-B)/(2MR^2)$ and the moment of inertia $I=(A+B+C)/3$, we have

\begin{equation}
	\label{eq:ci}
\frac{C}{MR^2}=\frac{I}{MR^2}+\frac{2}{3}J_2=0.37808.
\end{equation}

We can see from these data that the 2 parameters $\epsilon_1$ and $\epsilon_2$ can be straightforwardly derived, yielding:

\begin{eqnarray}
\epsilon_1 & = & 4.88230\times10^{-3}, \nonumber \\
\epsilon_2 & = & 2.92901\times10^{-3}. \nonumber
\end{eqnarray}

\par The 3 parameters related to the size and shape of the core depend on its composition. 
We consider 2 end-members: either the core is made of pure iron, its density $\rho_c$ being $8,000kg/m^3$, or
it is a eutectic mixture of FeS, yielding $\rho_c=5,150kg/m^3$ \citep{u1975}. Another uncertainty is in the composition of the 
crust. \citet{ajlms2001} considered 2 possibilities: either a thin $(\leq50km)$ crust with a low density 
$(\leq2,600kg/m^3)$, or a small amount of low density crust (that we propose to neglect) overlying a thicker (100-200 km) melt-rich asthenosphere 
(density between $3,000$ and $3,200 kg/m^3$). The reader can find additional information on Io's internal structure in \citet{msas2010}. With 
these assumptions, we get 6 interior models of Io (Tab.\ref{tab:param6int}).

\placetable{tab:param6int}

\begin{table}[ht]
\centering
\caption{Physical parameters of our 6 models. The models 1, 3 and 5 assume a pure iron core, while we have a eutectic FeS core in the models 2, 4 and 6. The models 3 and 4 
also consider a 30 km-thick crust, while the models 5 and 6 have a 150 km-thick crust. $\rho_s$ is the density of the crust, 
and $\rho_m$ the one of the mantle.\label{tab:param6int}}
\begin{tabular}{r|rrrrrr}
Models & $\rho_c$ & $\rho_m$ & $\rho_s$ & $\epsilon_3$ & $\epsilon_4$ & $\delta$ \\
 & $(kg/m^3)$ & $(kg/m^3)$ & $(kg/m^3)$ & $\times100$ & $\times100$ & $\times100$ \\
\hline
 1 & $8,000$ & $3,291$ & --      & $1.722$ & $2.210$ & $1.654$ \\
 3 & $8,000$ & $3,377$ & $2,400$ & $1.836$ & $2.673$ & $1.270$ \\
 5 & $8,000$ & $3,409$ & $3,100$ & $1.877$ & $2.819$ & $1.188$ \\
\hline
 2 & $5,150$ & $3,243$ & --      & $0.980$ & $1.057$ & $6.503$ \\
 4 & $5,150$ & $3,337$ & $2,400$ & $1.029$ & $1.243$ & $5.175$ \\
 6 & $5,150$ & $3,351$ & $3,100$ & $1.033$ & $1.253$ & $5.160$ \\
\hline
\end{tabular}
\end{table}

\par We can see that the 6 models can be splitted into 2 groups, the discrimination coming from the composition of the core.
The equations used to derive these models can be found in \citet{nkr2011}, Eq.1 to 12.

\section{Rotational dynamics}

\par As already said, we use the Poincar\'e-Hough model to represent the rotational dynamics of Io. This model considers a rigid outer layer, composed
of the mantle and eventually the crust, and a cavity filled by an inviscid fluid, constituting the fluid core. This cavity is triaxial, which allows 
pressure coupling at the core-mantle boundary. The computation of the rotational dynamics consists in numerically integrating the Hamilton equations
derived from the following Hamiltonian:

\begin{equation}
\label{equ:hamiltout}
\mathcal{H}(p,P,r,R,\xi_1,\eta_1,\xi_2,\eta_2)=\mathcal{H}_1(P,\xi_1,\eta_1,\xi_2,\eta_2)+\mathcal{H}_2(p,P,r,R,\xi_1,\eta_1),
\end{equation}
where $\mathcal{H}_1$ is the kinetic energy of the system, and $\mathcal{H}_2$ the perturbing potential of Jupiter.

We have (see e.g. \citet{n2012})

\begin{eqnarray}
\mathcal{H}_1&=&\frac{n}{2(1-\delta)}\Bigg(P^2+\frac{P_c^2}{\delta}+2\sqrt{\Big(P-\frac{\xi_1^2+\eta_1^2}{4}\Big)\Big(P_c-\frac{\xi_2^2+\eta_2^2}{4}\Big)}\big(\eta_1\eta_2-\xi_1\xi_2\big) \nonumber \\
 & & +2\Big(P-\frac{\xi_1^2+\eta_1^2}{2}\Big)\Big(\frac{\xi_2^2+\eta_2^2}{2}-P_c\Big)\Bigg) \nonumber \\
& & \nonumber \\
&+&\frac{n\epsilon_1}{2(1-\delta)^2}\Bigg(P_c^2-\Big(\frac{\xi_2^2+\eta_2^2}{2}-P_c\Big)^2+P^2-\Big(P-\frac{\xi_1^2+\eta_1^2}{2}\Big)^2 \nonumber \\
&+& 2\sqrt{\Big(P-\frac{\xi_1^2+\eta_1^2}{4}\Big)\Big(P_c-\frac{\xi_2^2+\eta_2^2}{4}\Big)}\big(\eta_1\eta_2-\xi_1\xi_2\big)\Bigg) \nonumber \\
& & \nonumber \\
&+&\frac{n\epsilon_2}{2(1-\delta)^2}\Bigg(\frac{1}{4}\big(4P-\xi_1^2-\eta_1^2\big)\big(\xi_1^2-\eta_1^2\big)+\frac{1}{4}\big(4P_c-\xi_2^2-\eta_2^2\big)\big(\xi_2^2-\eta_2^2\big) \nonumber \\
&-&2\sqrt{\Big(P-\frac{\xi_1^2+\eta_1^2}{4}\Big)\Big(P_c-\frac{\xi_2^2+\eta_2^2}{4}\Big)}\big(\eta_1\eta_2+\xi_1\xi_2\big)\Bigg) \nonumber \\
& & \nonumber \\
&-&\frac{n\epsilon_3}{2(1-\delta)^2}\Bigg(\delta \Big(P^2-\Big(P-\frac{\xi_1^2+\eta_1^2}{2}\Big)^2\Big)+\Big(P_c^2-(\frac{\xi_2^2+\eta_2^2}{2}-P_c\Big)^2\Big)\Big(2-\frac{1}{\delta}\Big) \nonumber \\
&+& 2\delta\sqrt{\Big(P-\frac{\xi_1^2+\eta_1^2}{4}\Big)\Big(P_c-\frac{\xi_2^2+\eta_2^2}{4}\Big)}\big(\eta_1\eta_2-\xi_1\xi_2\big)\Bigg) \nonumber \\
& & \nonumber \\
&+&\frac{n\epsilon_4}{2(1-\delta)^2}\Bigg(\frac{\delta}{4}\big(4P-\xi_1^2-\eta_1^2\big)\big(\eta_1^2-\xi_1^2\big)+\Big(2-\frac{1}{\delta}\Big) \frac{1}{4}\big(4P_c-\xi_2^2-\eta_2^2\big)\big(\eta_2^2-\xi_2^2\big)  \nonumber \\
&+ & 2\delta\sqrt{\Big(P-\frac{\xi_1^2+\eta_1^2}{4}\Big)\Big(P_c-\frac{\xi_2^2+\eta_2^2}{4}\Big)}\big(\eta_1\eta_2+\xi_1\xi_2\big)\Bigg), \label{equ:HG3}
\end{eqnarray}
and

\begin{equation}
\mathcal{H}_2(p,P,r,R,\xi_1,\eta_1)=-\frac{3}{2}\frac{\mathcal{G}M_{\jupiter}}{nd^3}\big(\epsilon_1(x^2+y^2)+\epsilon_2(x^2-y^2)\big),
\label{equ:pull2}
\end{equation}
where $\mathcal{G}$ is the gravitational constant, $M_{\jupiter}$ the mass of Jupiter, $n$ the orbital frequency of Io, and $d$ the distance Io-Jupiter. $x$, $y$ and 
$z$ are the coordinates of the vector pointing to Jupiter in the reference frame $(\vec{f_1},\vec{f_2},\vec{f_3})$ linked to the principal axes of inertia of Io.

The canonical variables of the problem are:

\begin{equation}
\begin{array}{lll}
p, & \hspace{2cm} & P=\frac{G}{nC}, \\
r, & \hspace{2cm} & R=P(1-\cos K), \\
\xi_1=-\sqrt{2P(1-\cos J)}\sin l, & \hspace{2cm} & \eta_1=\sqrt{2P(1-\cos J)}\cos l, \\
\xi_2=\sqrt{2P_c(1+\cos J_c)}\sin l_c, & \hspace{2cm} & \eta_2=\sqrt{2P_c(1+\cos J_c)}\cos l_c. \\
\end{array} \\
\end{equation}
Each of these 4 lines is related to a dynamical degree of freedom. The first one is related to the longitudinal motion of the whole body, $p$ is very close to the spin angle and $\vec{G}$ is the angular momentum of Io. $K$ is the obliquity with respect to the normal to the reference plane (here the equatorial plane of Jupiter at J2000.0), and $r$ is the node associated. The third degree of freedom is related to the polar motion (or wobble) of Io, $J$ being its amplitude. And the last one is related to the orientation of the velocity field of the fluid filling the core, we have in particular $P_c=G_c/(nC)$ where $\vec{G_c}$ is the angular momentum of the pseudo-core. The pseudo-core is very close to the core, it lacks of physical relevance but is convenient to write a Hamiltonian formulation of the equations. These canonical variables do not directly represent observables of the rotation (in fact only the surface can be observed), but exact observable quantities can be extracted from them. 

\par The coordinates of Jupiter come from real ephemerides of Io, here L1.2 \citep{ldv2006}. This way, we consider the orbital dynamics of Io with the most possible 
accuracy. The orbital period of Io is $\approx1.769$ day and its eccentricity $4\times10^{-3}$. It experiences a $462$-d periodic perturbation due to the proximity of the 
$2:1$ orbital resonance with Europa, and is locked into a laplacian orbital resonance involving also Europa and Ganymede.

\par The numerical integrations are performed with the 10th order Adams-Bashforth-Moulton predictor-corrector integrator. Once the solutions of the system have been obtained, we use
Laskar's NAFF algorithm \citep{l1993,l2005} to represent them as sums of quasiperiodic series, i.e.

\begin{equation}
\label{eq:xtn}
x(t)\approx\sum_{n=0}^{N}A_n^{\bullet}\exp\left(\imath\nu_n^{\bullet}t\right)
\end{equation}
for complex variables, or

\begin{equation}
 x(t)\approx\sum_{n=0}^{N} \mathcal{A}_n^{\bullet} \cos\left(\nu_n^{\bullet}t+\phi_n^{\bullet}\right)
\label{eq:naffrc}
\end{equation}
for real ones. In doing this, we can identify, in each variable of the problem, the influence of every single perturber (Jupiter, the other satellites, the Sun\ldots).

\par Io is assumed to be in a dynamical equilibrium known as Cassini State 1
\citep{c1693,c1966}, as a consequence the initial conditions of the numerical integrations should be appropriately chosen. Deriving the equilibrium related to a simplified system 
(e.g. one-dimensional rigid rotation and circular orbit of the perturber) is usually possible, but in a sophisticate system as we used, with complete ephemerides, it cannot be done
accurately enough without using a perturbation theory. It is possible to derive an approximate equilibrium, but the numerical solutions will exhibit some free librations around the equilibrium, that are
supposed to have been damped in the real system. For this reason, we chose to improve iteratively the initial conditions in using the algorithm by \citet{ndc2013}, consisting in

\begin{enumerate}
\item starting from "pretty acceptable" initial conditions,
\item running a numerical integration,
\item identifying the free librations around the equilibrium,
\item removing them from the initial condions, and reiterate the process.
\end{enumerate}
	
\par Once we have computed the equilibrum solution for our 6 models, we derive the following observable outputs: 

\begin{itemize}
\item longitudinal librations of the mantle $\phi_m=p_m-nt$, often called \emph{physical librations}, where $p_m$ is the spin angle of the mantle,
\item obliquity of the mantle $\epsilon_m$ (angle between the angular momentum of the mantle and the normal to the orbit of Io),
\item amplitude of the polar motion $J_m$,
\item tilt of the velocity field of the fluid $J_c$,
\end{itemize}
the formulae giving these quantities being extensively derived in \citet{ndl2010}. The results are gathered in Tab.\ref{tab:output}.

\placetable{tab:output}

\begin{table}[ht]
\centering
\caption{Variations of the outputs in the different models. $\phi_m$ is the amplitude of the 
longitudinal librations, $J_m$ the polar motion of the mantle, $J_c$ the tilt of the fluid, 
and $\epsilon_m$ is the obliquity of the mantle.\label{tab:output}}
\begin{tabular}{r|rrrrrr}
Models & $\phi_m$   & $\phi_m$   & $<J_m>$   & $<J_c>$    & $<\epsilon_m>$ & $\epsilon_m$ \\
       & ($462$ d)  & ($1.76$ d) &           &            &                & ($274$ d)    \\
       & (arcsec)   & (arcsec)   & (mas)     & (arcsec)   & (arcsec)       & (arcsec)     \\
\hline
 $1$ & $39.82462$   & $30.73077$ & $136.592$ &  $5.89884$ & $7.88443$      & $2.35306$    \\
 $3$ & $39.82475$   & $30.60912$ & $135.689$ &  $5.52936$ & $7.87640$      & $2.33113$    \\
 $5$ & $39.82475$   & $30.58332$ & $138.148$ &  $5.41307$ & $7.87537$      & $2.32659$    \\
\hline
 $2$ & $39.82303$   & $32.35491$ & $139.912$ & $15.82812$ & $8.14896$      & $3.60967$    \\
 $4$ & $39.82346$   & $31.89343$ & $140.550$ & $13.13468$ & $8.06166$      & $3.08910$    \\
 $6$ & $39.82345$   & $31.88809$ & $136.018$ & $12.78276$ & $8.04716$      & $3.07908$    \\
\end{tabular}
\end{table}

\par We can unfortunately see a high degeneracy in the sense that observing the rotation of Io should not allow to draw conclusions on its interior,
the differences between the outputs being too small. The only number changing significantly is the tilt of the velocity field of the fluid $J_c$, 
that cannot be directly observed.

\section{Elliptical instability}

\par In all the calculations, we have assumed that the flow of the fluid is laminar. In fact, as initially seen experimentally by \citet{p1986} and theoretically explained by 
\citet{b1986} in the context of an unbounced strained uniform vortex, the periodic forcing of the elliptical cavity on the underlying rotation state produces a pairwise 
resonance of inertial waves which can grow exponentially. These studies follow independent predictions in the 1970s, a recent review of the topic is given in \citet{k2002}. To 
check the stability of the flow we need to consider the growth rate $\sigma$ for an arbitrary perturbation $\vec{v}$, see e.g. \citet{km1998}:

\begin{equation}
\label{eq:km1998}
\sigma(t)=\frac{1}{2}\frac{d\ln<\vec{v}^2/2>}{dt}=-\frac{<\vec{v}\cdot\vec{\nabla} U\cdot\vec{v}>}{<\vec{v}^2>},
\end{equation}
where the flow $\vec{U}$ is the velocity field of the fluid a priori assumed to be laminar, and $<>=\int dV$, $V$ being the volume of the 
fluid. The flow is stable when $\sigma<0$, and unstable otherwise. \citet{clml2012} have recently derived the following formula:

\begin{equation}
\label{eq:cebron2012}
\sigma=n\left(\frac{17}{64}\epsilon\beta-2.62(1-\eta)\frac{1+\eta^4}{1-\eta^5}\sqrt{E}-\frac{\Lambda}{16}\right),
\end{equation}
in the context of a triaxial body perturbed by a primary whose rotation axis is close to the geometrical polar axis, and is under the influence of a magnetic field. The 
parameters involved in this formula are:

\begin{itemize}

\item $\epsilon$: amplitude of the physical librations (Tab.\ref{tab:output}) at the orbital frequency (period: 1.76 d)

\item $\beta=\frac{a_c^2-b_c^2}{a_c^2+b_c^2}$

\item $\eta$: ratio between the internal and the external radii of the fluid layer. It is equal to $0$ for fully liquid cores.

\item Ekman number $E=\frac{\nu}{\Omega R_c^2}$ where $\nu$ is the fluid kinematic viscosity, and $\Omega=nP_c/\delta=n$ the mean velocity of the fluid. This is a kind of adimensional viscosity.

\item Elsasser number $\Lambda=\frac{\sigma_eB_0^2}{\rho_c\Omega}$, where $B_0$ is the intensity of the magnetic field and $\sigma_e$ the fluid electrical 
	conductivity. This contribution of the magnetic field has been derived by \citet{clml2012}, generalizing a result by \citet{hll2009}.

\end{itemize}
As \citet{hll2009} and \citet{clml2012} did, we took $B_0=1850nT$, $\nu=10^{-6}m^2s^{-1}$ and $\sigma_e=4\times10^5S.m^{-1}$. The results are given in Tab.\ref{tab:instabresult}.

\placetable{tab:instabresult}

\begin{table}[ht]
\centering
\caption{Grow rate $\sigma$, and the time associated compared with the spin-up time $t_{spin-up}$.\label{tab:instabresult}}
\begin{tabular}{r|rrr}
Models & $\sigma$ & $1/\sigma$ & $t_{spin-up}$ \\
 & $(yr^{-1})$ & $(kyr)$ & $(kyr)$ \\
\hline
 $1$ & $1.142\times10^{-3}$ &  $0.875$ & $3.321$ \\
 $3$ & $1.561\times10^{-3}$ &  $0.640$ & $3.140$ \\
 $5$ & $1.700\times10^{-3}$ &  $0.588$ & $3.109$ \\
\hline
 $2$ & $3.325\times10^{-4}$ &  $3.008$ & $4.770$ \\
 $4$ & $2.260\times10^{-4}$ &  $4.425$ & $4.559$ \\
 $6$ & $2.353\times10^{-4}$ &  $4.250$ & $4.557$ \\
\end{tabular}
\end{table}

\par The mean growth rate $\sigma$ is positive in every model. Anyway, it is smaller in the "even" models, i.e. with a FeS core, with a growth time of the order of $3,000$ years, while it is always smaller than $1,000$ years in the "odd" models (Fe core). For comparison we give in the last column the 
spin-up time $t_{spin-up}=1/(n\sqrt{E})$ \citep{gh1963}. This is the typical spin-up/spin-down time necessary for the fluid to recover the mantle velocity. $t_{spin-up}$ is 
usually assumed to be long enough so that the velocity of the fluid can be considered as constant. We can notice that it is of the same order of magnitude as the growth time $1/\sigma$.

\par In all these models we find that the inertial waves of the core of Io should be unstable. The work of \citet{km1998} showed that the waves are unstable, and \citet{hll2009} confirmed the result even when the effect of the magnetic field is included. This last reference suggests a growth time of $\approx63$ years. However, \citet{clml2012} argue that 
this growth rate has been calculated in an extreme optimal case, in particular in considering an extremum of the instantaneous departure from the synchronous rotation. They 
conclude that the inertial waves in Io's fluid core should be stable. Our result lies between these two opposite conclusions, i.e. a positive but quite small growth rate. The main reason why our result is different from Cebron's with the same formula is that our $\beta$ (equatorial ellipticity of the core) is bigger. They also assumed a smaller amplitude of libration, derived by \citet{cb2003} for a rigid Io.

\section{Energy budget}

\par Io is known for its energy dissipation resulting in volcanos at its surface. Several attempts have been made to quantify this dissipation. One way to proceed is to 
use observations of Io's surface:

\begin{itemize}

\item \citet{vmjbg1994} estimated the global heat flow $\phi$ to be bigger than $2.5W/m^2$,

\item \citet{mjvbd2001} estimated it to be smaller than $13W/m^2$,

\item \citet{rstmbt2004} estimated $\phi$ to be between $2.0$ and $2.6W/m^2$, by studying several hot spots,

\end{itemize}
the reader can find additional references in \citet{vdmjwr2012}, Tab.2. Another way is to try to detect the influence of tidal dissipation on the orbit of Io by comparing astrometric 
observations with dynamical models. Recently, \citet{lakv2009} estimated the tidal dissipation to be $\dot{E}=(9.33\pm1.84)\times10^{13}$ W, yielding $\phi=2.24\pm0.45$ $W/m^2$ assuming 
that energy is transported out of Io at the same rate.

\par The energy dissipation due to the instability of the fluid core is \citep{llllr2010}:

\begin{eqnarray}
\dot{E} & = & -\frac{8\pi}{3}\rho_cR_c^4\sqrt{\nu}\left|\Omega^s-\Omega^o\right|^{5/2}\left|\omega_{SO}\right|^{5/2} \label{eq:lebars2}
\end{eqnarray}
where $M_c$ is the mass of the fluid core, $R_c$ its radius, $\rho_c$ its density, $\Omega^s=nP$ the spin rate of the mantle, and $\omega_{SO}$ is a normalized 
frequency of the spin-over mode, appearing in case of instability. We have $\omega_{SO}=\Omega^f/\Omega^s=1/P$, since the spin rate of the fluid is assumed to be $n$.

\par Using formula (\ref{eq:lebars2}) and $|\Omega^s-\Omega^o|=2e|\cos nt|$ where $e\approx0.004$ is the orbital eccentricity of Io, we get a mean $\dot{E}$ of $8.47\times10^8$ W for 
Model 1 and $2.32\times10^9$ W for Model 2. This is consistent with the conclusions of \citet{llllr2010} announcing a peak dissipation of $\approx4\times10^9$ W. This quantity is 
very small with respect to the energy dissipated by tides, i.e. $(9.33\pm1.84)\times10^{13}$ W.

\par Another possible source of dissipation is radioactive energy, because of the decay of unstable isotopes in the rock component of the body. \citet{hclmstv2010} estimate it between 
$3.08\times10^{11}$ W and $5.14\times10^{11}$ W depending on the composition of Io.

\placetable{tab:energy}

\begin{table}[ht]
\centering
\caption{Energy budget of Io, at the present time.\label{tab:energy}}
\begin{tabular}{lrr}
	Source of energy & $\dot{E}$ & Reference \\
\hline
Surface heat flow & $8.33-10.83\times10^{13}$ W & \citet{rstmbt2004} \\
Tidal dissipation & $(9.33\pm1.84)\times10^{13}$ W & \citet{lakv2009} \\
Radioactive energy & $3.08-5.14\times10^{11}$ W & \citet{hclmstv2010} \\
Fluid instability & $8.47\times10^8-2.32\times10^9$ W & this note\\
\end{tabular}
\end{table}

\par The results are gathered in Tab.\ref{tab:energy}. We can see that if the flow of the fluid constituting the core of Io is unstable, the energy involved 
is negligible with respect to the tidal dissipation, as already stated \citet{llllr2010}. So, this cannot be responsible for Io's volcanic activity.

\section{Conclusion}

\par In this study we used the Poincar\'e-Hough model to predict the rotation of Io, depending on its internal structure. For that, we elaborated 6 interior
models, in considering pure iron or FeS compositions of the core, and the presence or not of a different crust. This rotational model describes the behavior
of a 2-layer body composed of a rigid mantle and a fully liquid core, in considering pressure coupling at the core-mantle boundary.

\par For each case considered, the amplitude of the short longitudinal librations is about $30$ arcsec, and the mean obliquity $\approx8$ arcsec, these 
two quantities being a little bigger for a eutectic FeS core, which is larger. But the main difference is in the tilt of the angular momentum 
of the fluid constituting the core, that is bigger (between $10$ and $15$ arcsec) for the FeS than for the pure Fe one ($\approx5$ arcsec). These 
differences should unfortunately not be detectable. A study of the elliptical instability indicates that the inertial waves should be unstable 
in any case, the growth time being between $500$ and $800$ yr for a Fe core and between $3$ and $5$ kyr for a FeS core. This last calculation only 
considers the influence of the longitudinal librations, without involving the tilt of the fluid. This instability does not have a significant 
impact on the energy dissipated at the surface.

\section*{Acknowledgements}

This research used resources of the Interuniversity Scientific Computing Facility located at the University of Namur, Belgium, which is supported by the F.R.S.-FNRS under 
convention No. 2.4617.07. The author is F.R.S.-FNRS post-doctoral research fellow, and is indebted to David C\'ebron for fruitful discussions.





\end{document}